\documentclass{iopart}
\usepackage{amssymb}
\usepackage{graphicx}
\usepackage{verbatim}
\usepackage[colorlinks=true,linkcolor=blue,citecolor=blue,urlcolor=blue]{hyperref}
\usepackage{txfonts}

\begin{document}

\title{Nonlinear Spectroscopy of Controllable Many-Body Quantum Systems}
\author{Manuel Gessner$^{1,2}$, Frank Schlawin$^{1,3}$, Hartmut H\"affner$^{2}$, Shaul Mukamel$^{3}$ and Andreas Buchleitner$^1$}
\address{$^1$ Physikalisches Institut, Albert-Ludwigs-Universit\"at Freiburg, Hermann-Herder-Stra\ss e 3, 79104 Freiburg, Germany}
\address{$^2$ Department of Physics, University of California, Berkeley, California 94720, USA}
\address{$^3$ Department of Chemistry, University of California, Irvine, California 92697, USA}
\ead{\mailto{manuel.gessner@physik.uni-freiburg.de}, \mailto{frank.schlawin@physik.uni-freiburg.de}}
\date{\today}

\begin{abstract}
We establish a novel approach to probing spatially resolved multi-time correlation functions of interacting many-body systems, with scalable experimental overhead. Specifically, designing nonlinear measurement protocols for multidimensional spectra in a chain of trapped ions with single-site addressability enables us, e.g., to distinguish coherent from incoherent transport processes, to quantify potential anharmonicities, and to identify decoherence-free subspaces.
\end{abstract}

\maketitle

\section{Introduction}

To unravel the spectral and dynamical properties of composite quantum systems of increasing complexity constitutes an indispensable prerequisite for robust control in diverse areas of modern quantum science, from quantum information processing \cite{HARTMUTREVIEW,Walter,Sciarrino} over photo-induced chemical reactions \cite{Brixner} to the primary processes of light-energy conversion in nature and technology \cite{Rienk, Durrant}. A key ingredient of truly complex quantum systems are strong correlations over broadly distributed energy and time scales \cite{Pfeiffer}, which may lead to critical behavior or emergent phenomena, and become manifest in such systems' dynamics \cite{AbuChaosReview,Anderson}.

For a precise characterization of the latter, higher-order correlation functions need to be probed, which poses a formidable experimental challenge \cite{Ernst,HammZanni,Shaul_book}. Much progress has been accomplished in this regard in ultrafast multidimensional spectroscopy of molecular aggregates, where experiments are often carried out under extremely challenging conditions characterized by short timescales and tight spatial confinement \cite{Shaul_book,Fleming,Kauffmann,Engel}. In cold matter experiments, however, where complexity is constructed by controlled assembly of individual components \cite{AbuChaosReview,CHRISTIAN,QuantumGasesRMP,RydbergReview,John,Plenio13}, Ramsey-type pump-probe techniques are still the preferred experimental tool to probe the many-body dynamics \cite{Lukin13}.

In a Ramsey experiment, the system evolves through a coherent superposition of ground- and excited state in a two-pulse sequence. Such experiments can resolve the energies of contributing excited states, as well as decoherence rates caused by environmental couplings; they are however restricted in their combined frequency-time resolution \cite{Shaul_book}, and cannot resolve manifolds of higher excited states or the contribution of individual pathways. For this reason, in chemical physics one typically employs four-pulse sequences combined with a phase-dependent selection of quantum pathways to resolve involved transport processes, exciton couplings, and conformational changes in large molecular aggregates, such as photosynthetic complexes.

In this Paper we combine this systematic construction of measurement protocols of higher-order correlation functions with the extraordinary control and, in particular, the single-site addressability of cold-matter systems \cite{Markus10,Weitenberg11,HARTMUTREVIEW}. We elaborate how our general formalism provides unmatched possibilities for the direct experimental assessment of spatially resolved multi-time correlation functions, in experiments which rely on current trapped-ion technology. In particular, second- and fourth-order signals will be defined to monitor intricate dynamical features in the vibrational and electronic degrees of freedom.

\section{Nonlinear Measurement Protocols}
Inspired by multidimensional optical spectroscopy, we develop a general formalism to systematically construct nonlinear measurement protocols in quantum many-body systems using a diagrammatic theory. To provide physical intuition, we consider the example of vibrational degrees freedom in a chain of $N$ ions in a linear trap. However, the obtained theoretical formalism is completely general and independent of the system or degree of freedom. 

The ion trap potential can be characterized by trap frequencies $(\nu_x,\nu_y,\nu_z)$, with the confinement much stronger in the transverse than in the axial direction ($\beta:=\nu_z^2/\nu_{x}^2 \ll 1$) \footnote{We assume that the harmonic confinement along the two uncoupled transverse directions is of the same order of magnitude, i.e., $\nu_x\approx\nu_y$ \cite{PorrasCirac}.}. Transverse vibrations along the $x$-direction can then be described by a tight-binding Hamiltonian of local phonons including a tunneling term \cite{PorrasCirac}, 
\begin{equation}\label{eq.phononhamiltonian}
H=\sum_{i=1}^N\omega^0_ia^{\dagger}_ia_i+\sum_{i<j}t_{ij}(a^{\dagger}_ia_j+a^{\dagger}_ja_i),
\end{equation}
where $a^{\dagger}_i$ denotes the creation operator for a local phonon at site $i$, $a^{\dagger}_i|0\rangle = \vert 1_i \rangle$. The local trap frequencies and the coupling matrix can be microscopically derived as \cite{PorrasCirac}
\begin{eqnarray} \label{eq.phonon-site-energy}
\omega^0_{i}/\nu_x&=1-\frac{\beta}{2}\sum_{j\neq i}\frac{1}{|u^0_i-u^0_{j}|^3},\\
t_{ij}/\nu_x&=\frac{\beta}{2}\frac{1}{|u^0_i-u^0_{j}|^3},
\end{eqnarray}
with $u^0_i=z^0_i/l_0$, where $z_i^0$ denote the ion's equilibrium positions, and a typical length scale is given by $l_0^3=e^2/(m\nu_z^2)$ \cite{James}. An additional anharmonic potential $U\sum_ia^{\dagger 2}_ia_i^2$ with tunable strength $U$ can be induced by a standing electromagnetic wave, effectively generating a Bose-Hubbard model with long-range couplings \cite{PorrasCirac}.

The basic building blocks for multidimensional spectroscopic protocols are excitation and readout schemes. A vibrational excitation of the form $D_j(\alpha e^{i\phi})\approx \mathbb{I} + \alpha e^{i\phi}a_j^{\dagger} - \alpha e^{-i\phi}a_j$ can be generated by suitable, perturbative pulse sequences \cite{Brown,Harlander}, where $\alpha\ll 1$, and $\phi$ can be controlled through the pulse parameters. The superoperator $\mathcal{V}_j(\phi)[\rho] = D_j(\alpha e^{i\phi})\rho D^{\dagger}_j(\alpha e^{i\phi})$ describes the corresponding change of the density matrix induced by this weak interaction. In the course of this paper, we consider the impulsive limit for the interaction $\mathcal{V}$, which means that the duration of the light pulses is much shorter than the time scale of the vibrational dynamics. This is justified since the characteristic time scale of the Hamiltonian evolution can be adjusted by controlling the distance between neighboring ions via the parameter $\beta$ \cite{PorrasCirac}. Excitations of local phonons can be read out via the observable $A_j  = \sum_{n_j = 1}^{\infty} \sin^2 \left( \sqrt{n_j} \pi/2 \right) \vert n_j \rangle  \langle n_j  \vert$ ($j$ labels the measured ion), which is accessible by mapping the vibrational population onto the ion's electronic state \cite{Heinzen} {(see \ref{sec.A}).

Using these ingredients, multidimensional spectra are obtained by scanning the time-delays between a sequence of pulses and detecting the observable $A_j$. After a series of $m$ interactions on the ions $i_1,\dots,i_m$, the signal is given by 
\begin{equation}
S^{(m)}_{i_1,\dots,i_m;j}(t_1,\dots,t_m)= \big\langle A_j\big\rangle = \mathrm{Tr}\{A_j \rho_{i_1,\dots,i_m}^{(m)}(t_1,\dots,t_m)\},
\end{equation}
where the expectation value is taken with respect to the non-equilibrium density matrix created by a succession of short, impulsive interaction events and unperturbed time evolution,
\begin{equation}
\rho_{i_1,\dots,i_m}^{(m)}(t_1,\dots,t_m) = \mathcal{G}(t_m)\mathcal{V}_{i_m}(\phi_m)\dots \mathcal{G}(t_1)\mathcal{V}_{i_1}(\phi_1) [\rho(0)]. \label{rho^m}
\end{equation}
The latter is described by the Green's function $\mathcal{G}(t)=\exp \mathcal{L}t$, where $\mathcal{L}$ denotes the Liouville superoperator \cite{Shaul_book,BREUERBOOK}, which we assume time-independent: $\mathcal{L}[\rho] = -i[H,\rho] + \sum_i(L_i\rho L_i^{\dagger}-\frac{1}{2}\{L_i^{\dagger}L_i,\rho\})$. Different Lindblad operators $L_i$ \cite{BREUERBOOK} can be simulated in trapped-ion systems \cite{Poyatos,Myatt,Julio}.

These signals constitute a generalization of Ramsey spectroscopy, and provide the possibility to distinguish the contributions of individual quantum pathways. To see this, note that each time an excitation is created or destroyed due to the interaction $\mathcal{V}_{i_k}(\phi_k)$, the phase $\phi_k$ of the $k$-th pulse is imprinted on the quantum state. Hence, different contributions to the total signal can be distinguished by their dependence on these phases. One can exploit this to post-select signals pertaining to certain pathways with distinct combinations of coherences and populations: To this end one employs the phase cycling protocol originally developed in nuclear magnetic resonance \cite{Ernst}. It is implemented by repeating the experiment for a small discrete set of phases followed by Fourier analysis \cite{Ernst,Hulst}. Individual pathways can be represented by Feynman diagrams \cite{Shaul_book}, which offer an intuitive interpretation of the signals (see also figure~\ref{fig.2}(a)). Further information on coherent pathways and their selection via phase cycling can be found in the \ref{sec.pc}.

Let us stress that the way such signals are obtained in controlled quantum systems holds considerable advantages over related methods, such as phase-matched heterodyne ensemble spectroscopy of bulk materials \cite{HammZanni,Shaul_book} or phase-sensitive fluorescence measurements \cite{LottPNAS}, e.g., of single molecules \cite{Hulst}. Specifically, we point out three key differences between these methods and the approach proposed here. 

\begin{itemize}
\item The first consists in the ability to create localized excitations due to the micrometer separation of the ions, which induces superpositions of a multitude of eigenstates with a single interaction. This is not possible, for instance, when working on molecular aggregates, where interacting chromophores are separated only by few nanometers, two orders of magnitude below the diffraction limit of optical light. 

\item Second, since artificial quantum systems, as considered in the present manuscript, are typically well isolated from the environment, the induced excitations do not decay naturally within the relevant experimental time-scales \cite{HARTMUTREVIEW}. Instead, the described fluorescence signal must be induced externally via the coupling to a short-lived state, providing full control over the delay between readout and the final interaction. Direct field measurements of the Raman scattered light are not suitable for these single quantum systems due to the insufficient number of scattered photons. This is in stark contrast to fluorescence measurements in molecular systems, where fluorescence signals are created by spontaneous decay after a random time, and heterodyne measurements of stimulated emission, which is induced immediately following the last excitation. Therefore, methods which were developed for phase-matched heterodyne electronic spectroscopy cannot be implemented straight-forwardly and instead must be adjusted to the current fluorescence-based situation. However, as we will show later in this manuscript, analogs of well-known measurement protocols from electronic two-dimensional spectroscopy can be defined and interpreted similarly. 

\item Third, aside from strong spatial confinement, experiments on molecular aggregates are also characterized by extremely challenging timescales. For instance, typical timescales for energy transport and coherence decay on photosynthetic complexes are on the order of picoseconds \cite{Blankenship,Rienk}. These parameters render precise experimental control of such molecular aggregates extremely demanding. In contrast, quantum optical experiments, such as trapped ions, allow to study energy transport under well-controlled conditions on microsecond timescales with milliseconds of coherent dynamics \cite{Brown,Harlander}.
\end{itemize}

Let us also briefly comment on the difference of single-shot experiments on an ensemble of systems in a bulk and experiments on single quantum systems, which are repeated many times to obtain a statistically relevant sample of results. In the latter case, ensemble-averaging over slightly fluctuating conditions leads to decoherence effects whose phenomenology is similar to the one experienced in bulk materials. In ion trap experiments, those effects are especially relevant in the context of quantum computation, and consequently have been analyzed thoroughly \cite{Schindler}. For the application of the proposed methods, these effects do not pose a problem, as demonstrated, for instance, by the successful implementation of related methods with single molecules \cite{Hulst}.

Making use of the advantages discussed above, we can now design nonlinear signals to target specific dynamical and spectral features of engineered quantum matter systems. Current quantum optical experiments struggle with the efficient analysis of long-range coherences, which are responsible for notoriously elusive quantum critical phenomena \cite{Islam13}, and whose role is also currently debated in quantum transport processes \cite{Engel} which in turn may be studied in quantum simulations \cite{Plenio13}. Moreover, in the light of increasing complexity, appropriate methods to certify and diagnose quantum computations are lacking -- in particular for the unambiguous identification of sources of error, such as anharmonic corrections to the trap potential and the precise characterization of detrimental effects induced by external environments \cite{Schindler}. In the following, we show how nonlinear spectroscopy is able to improve on this situation, by demonstrating a selection of applications for examples of second- and fourth order signals.

\section{Applications}
\subsection{Coherent and incoherent phonon transport}

\begin{figure*}[tb]
\centering
\includegraphics[width=.99\textwidth]{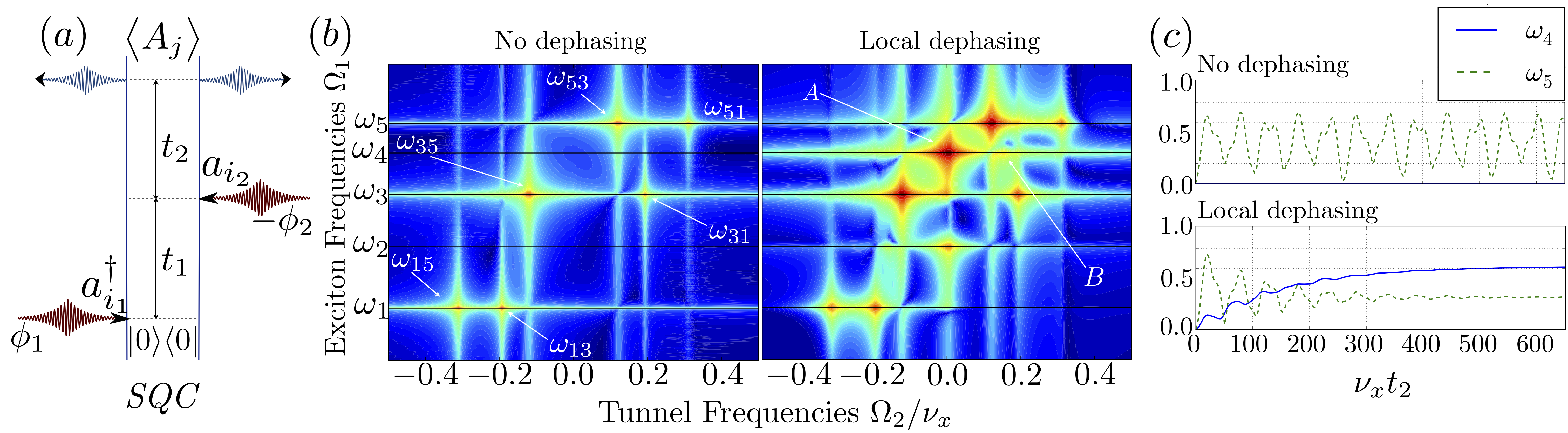}
\caption{The single quantum coherence (\textit{SQC}) signal unveils space- and time-resolved exciton evolution. (a) The 
Feynman diagram describing $S_{i_1,i_2;j}^{\left(\textit{SQC}\right)}$ depicts those contributions to the impulsive 
excitations (red arrows) $\mathcal{V}_j(\phi)$, which imprint the 
phase label $\phi_1-\phi_2$. Blue arrows indicate read-out (see \ref{sec.pc} for an interpretation of the diagrams).
(b) $|S_{1, 1; 3}^{\left(\textit{SQC}\right)} (\Omega_1, \Omega_2)|$ reveals the contributing exciton spectrum (along $\Omega_1$) 
and tunnelling rates along $\Omega_2$, for unitary dynamics (left), and under additional local dephasing with strength $\gamma = 0.01\nu_{x}$ (right). The signals are rescaled by $\mathrm{arcsinh}$ to highlight small features. (c) Time-resolved exciton transport can be 
read off from $|S_{1, 1; 3}^{\left(\textit{SQC}\right)} (\Omega_1, t_2)|$ for $\Omega_1 = \omega_4$ (blue, solid line) and for $\Omega_1 = \omega_5$ (red, dashed line). The emergence of a 
signal at $\Omega_1=\omega_4$ under the influence of noise indicates incoherent exciton transport. At transient times, coherent exciton couplings can be observed as well. For $\beta=0.1$, diagonalization of the Hamiltonian (\ref{eq.phononhamiltonian}) yields $\omega_1=0.69\nu_{x}$ and $\omega_5=\nu_{x}$.}
\label{fig.2}
\end{figure*}

As a first application, we scrutinize the role of quantum coherence for phonon transport in an ion chain, via a second-order signal which monitors the space- and time-resolved spectral decomposition of a local excitation. To this end, we introduce the single quantum coherence (\textit{SQC}) signal,
\begin{equation}
S^{\left(\textit{SQC}\right)}_{ i_1, i_2; j}(t_1,t_2)=\mathrm{Tr}\{A_j\mathcal{G}(t_2)[\mathcal{G}(t_1)[a^{\dagger}_{i_1}\rho(0)]a_{i_2}]\}
\end{equation} 
which is extracted by phase cycling with respect to the phase $\phi_1 - \phi_2$ [Figure~\ref{fig.2} (a)] from the total second-order signal $S^{(2)}_{ i_1, i_2; j}(t_1,t_2)$. It contains time-correlation functions whose spatial resolution is controlled by the choice of excitation and readout pulses: The first tunable delay $t_1$ between the excitation pulses at ions $i_1$ and $i_2$ and a second tunable delay $t_2$ before readout at ion $j$ are scanned. To unveil spectral properties, we Fourier transform the signal with respect to one or both time delays, e.g., 
\begin{equation}
S^{\left(\textit{SQC}\right)}_{i_1, i_2; j} (\Omega_1, \Omega_2) = \int_0^{\infty} \!\! dt_1 \int_0^{\infty} \!\! dt_2 \; e^{i (\Omega_1 t_1 + \Omega_2 t_2)} S^{\left(\textit{SQC}\right)}_{i_1,i_2; j} (t_1, t_2).
\end{equation}

In figure~\ref{fig.2} we simulate \cite{qutip} \textit{SQC} signals for a chain of five ions initialized in their individual ground states $|0\rangle$ with confinement parameter $\beta = 0.1$. The first two pulses create vibrational excitations at the leftmost ion, i.e., $i_1 = i_2 = 1$, which are then probed, after interaction-induced propagation, at the central ion, $j = 3$. For the interpretation of the spectrum it is most instructive to represent the local excitations in terms of single exciton states (energy eigenstates of the chain) $|e_j\rangle$, e.g., $a^{\dagger}_{1}|0\rangle = \vert 1_1 \rangle = \sum_j c_{1j} \vert e_j \rangle$, which evolve with their respective eigenfrequencies $\omega_j$. We emphasize that the index $j$ now refers to an eigenstate of the chain, i.e., we now work in the exciton basis -- in contrast to Eq. (\ref{eq.phononhamiltonian}), where the Hamiltonian was described in a local site basis. Since the first pulse induces the coherence $|1\rangle\langle 0|$, which is selected via its phase signature $\phi_1$ [see figure~\ref{fig.2}(a)], the Fourier transform with respect to the first time delay $t_1$ reveals the single exciton frequencies $\omega_j$. During $t_2$, the contributions with the phase label $\phi_1-\phi_2$ involve coherences between distinct excitons, such that the Fourier transform of $t_2$ reveals the energy differences $\omega_{ij} = \omega_i - \omega_j$ [Figure~\ref{fig.2}(b)]. As can be easily seen, these determine the periods at which an excitation coherently tunnels between the ions: The probability for a phonon localized on ion $a$ to tunnel to ion $b$ at a time $t$ is given by
$p_{ab}(t) = |\langle 1_b \vert e^{- i H t} \vert 1_a \rangle|^2 = \sum_{ij} c_{ij} e^{- i \omega_{ij} t},$ with $c_{ij}=\langle 1_b | e_i \rangle \langle e_i \vert 1_a \rangle\langle 1_a | e_j \rangle \langle e_j \vert 1_b \rangle$. We will therefore denote the frequency differences $\omega_{ij}$ as tunnel frequencies. We remark here that the measurement protocol which reveals these frequencies only requires two localized excitation pulses and one readout pulse, \textit{independently} of the length of the chain.

In the right panel of figure~\ref{fig.2}(b), we now add local dephasing, described by $L_i=\sqrt{\gamma} a_i^{\dagger}a_i$ \cite{Myatt,Poyatos}. The coupling to the environment causes additional transitions, which induce incoherent transport. To understand this, we plot $|S_{1, 1; 3}^{\left(\textit{SQC}\right)} (\Omega_1, t_2)|$ in figure~\ref{fig.2}(c), which displays the evolution of the initial state's frequency components during $t_2$ and thereby allows to monitor time-resolved exciton transport. Without dephasing (top panel) there is no signal at $\omega_4$, since the associated 
breathing mode leaves the central ion immobile \cite{James}, and is therefore not detected by local readout. However, in the presence of local dephasing, the breathing mode is incoherently coupled to other modes, which do have a finite amplitude at the central ion. Thus, the bottom panel of figure~1(c) shows the emergence of a signal at $\omega_4$ with increasing $t_2$. We observe progressively incoherent (non-oscillatory) transport, leading to a pronounced two-dimensional signal at $(\Omega_1,\Omega_2)=(\omega_4,0)$ [peak A]. The transient oscillations on top are due to short-lived coherences between the excitons $|e_4\rangle$ and $|e_2\rangle$, and give rise to another, weak two-dimensional signal at $(\omega_4,\omega_{42})$ [peak B]. Thus, our local read-out scheme provides full information on spatially resolved transport and on the coherently or incoherently populated transporting eigenmodes of the chain.

\subsection{Detection of anharmonicities}

\begin{figure*}
\centering
\includegraphics[width=.99\textwidth]{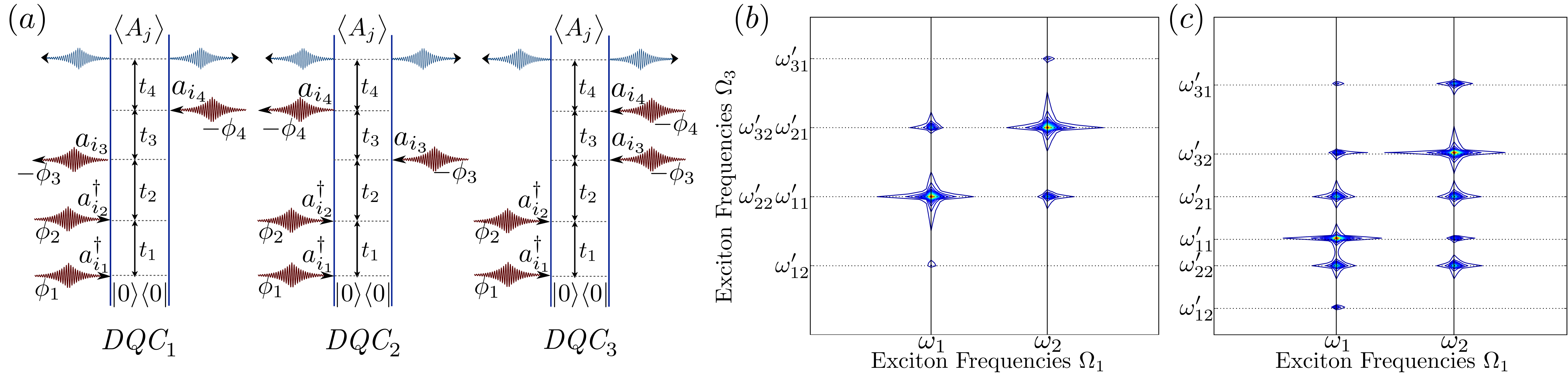}
\caption{The double quantum coherence (\textit{DQC}) signal is ideally suited for a spectral characterization of trap anharmonicities and the identification of excitation pathways to the double exciton manifold. (a) The three diagrams with phase signature $\phi_1 + \phi_2 - \phi_3 - \phi_4$ constitute the \textit{DQC} signal. (b) In a harmonic trap ($U=0$), $|S^{\left(\textit{SQC}\right)}_{0000; 0} (\Omega_1, t_2 = 0, \Omega_3, t_4=0)|$ displays degenerate double excitons. (c) A symmetry-breaking anharmonic correction ($U = - 0.025 \nu_{x}$) lifts these degeneracies and creates new pathways, as revealed, e.g., by the peak at $(\omega_1,\omega'_{31})$. The $\Omega_3$ axis is the same in plots (b) and (c), however the resonances are shifted in panel (c), due to the anharmonicity. Both plots simulate signals of two ions with $\beta=0.1$, leading to $\omega_1=0.95\nu_{x}$ and $\omega_2=\nu_{x}$.}
\label{fig.3}
\end{figure*}

The second order signal employed in our example above can only probe excitonic states within the single excitation manifold (due to the very structure of the impulsive interaction $\mathcal{V}_j(\phi)$), and is therefore unsuitable to detect any anharmoniticity of the phonon spectrum. For the latter purpose, we need to probe the {\em double} excitation manifold, and, hence, to find an appropriate fourth-order signal. Such an observable is given by the double quantum coherence (\textit{DQC}), which has been successfully employed to probe electronic correlations in molecules \cite{Scholes-review} (see also \cite{Nemeth}). For our present purposes, we define  the analogous fluorescence-based signal, consisting of the four-pulse pathways with phase signature $\phi_1 + \phi_2 - \phi_3 - \phi_4$ [Figure~\ref{fig.3}(a)]. The first two pulses create a double exciton (two-phonon) state $|f_i\rangle$, which is subsequently probed by the third and fourth pulse. During $t_1$, the created coherence evolves with frequencies $\omega_i$, whereas during $t_3$, it oscillates with either $\omega_i$ ($\textit{DQC}_1$) or $\omega'_{ij}=\omega_{f_i}-\omega_{j}$ ($\textit{DQC}_2$ and $\textit{DQC}_3$), where $\omega_{f_i}$ are the eigenfrequencies of the double exciton states $|f_i\rangle$. Fourier transform with respect to the time delays $t_1$ and $t_3$ reveals signal contributions at $\omega_i$ and $\omega'_{ij}$, along $\Omega_1$ and $\Omega_3$, respectively. In the following, the single and double exciton energies $\omega_i$ and $\omega_{f_i}$ are labeled in ascending order, i.e. $\omega_1<\omega_2<\ldots$, and $\omega_{f_1}<\omega_{f_2}<\ldots$, respectively.

Let us consider the example $N=2$, where all pulses are applied to the same ion and the times $t_2$ and $t_4$ are kept zero. Note that, as pointed out in the introduction, in contrast to other methods, this approach allows to scan the time interval $t_4$ as well, which can be extremely useful for studies of, for instance, population decay. For the detection of anharmonicities, however, only the time intervals $t_1$ and $t_3$ need to be considered. Since for vanishing anharmonicity, $U=0$, the energy of each double exciton state $|f_i\rangle$ is given by the sum of two single exciton energies, $\omega_{f_i} = \omega_j + \omega_k$, transitions such as $\omega'_{11}$ and $\omega'_{22}$ coincide in figure~\ref{fig.3}(b). Moreover, peaks involving $\omega_{f_j}=2\omega_i$ may only be observed along $\Omega_1 = \omega_i$ since only if the first pulse excites $|e_i\rangle$, the second pulse can promote the system into $|f_j\rangle$. For instance, we do not observe a peak at $(\omega_1,\omega'_{31})$, since $\omega_{f_3}=2\omega_2$. 

When we add anharmonicity to the system ($U\neq 0$), two effects can be observed [Figure~\ref{fig.3}(c)]: First, degeneracies are lifted, and the signals at $\omega'_{1 1}$ and $\omega'_{2 2}$ can be resolved. Second, the anharmonicity $U$ perturbs the symmetry of double excitons. This opens up new excitation pathways, e.g., $|f_3\rangle$ can be accessed via $|e_1\rangle$, creating a signal at $(\omega_1, \omega'_{3 1})$. The strength of the anharmonicity $U$ can be inferred by comparing the distance of the $\omega'_{ij}$-transitions to the possible harmonic transitions, which in turn can be recovered from the single-exciton frequencies along the $\Omega_1$-axis.

\subsection{Two-dimensional lineshapes}
So far, we have demonstrated how to assess spectral properties of the \textit{vibrational} degree of freedom of the ion chain. However, each ion also carries an \textit{electronic} degree of freedom, which can be treated as a spin-1/2 system. The vibrational modes then mediate the coupling between the different spins. Applications of this type of interactions for quantum information purposes potentially suffer from the detrimental influence of environmental noise, what renders a precise characterization of decoherence mechanisms highly desirable in any such experiment. In our last example to illustrate the versatility of nonlinear spectroscopic techniques in combination with single-site resolution, we demonstrate how the line shape of a two-dimensional signal can reveal the nature and strength of the environmental coupling and identify a decoherence-free subspace. We further show how this can be used to certify the fidelity of Bell-state generation.

\begin{figure}
\centering
\includegraphics[width=.79\textwidth]{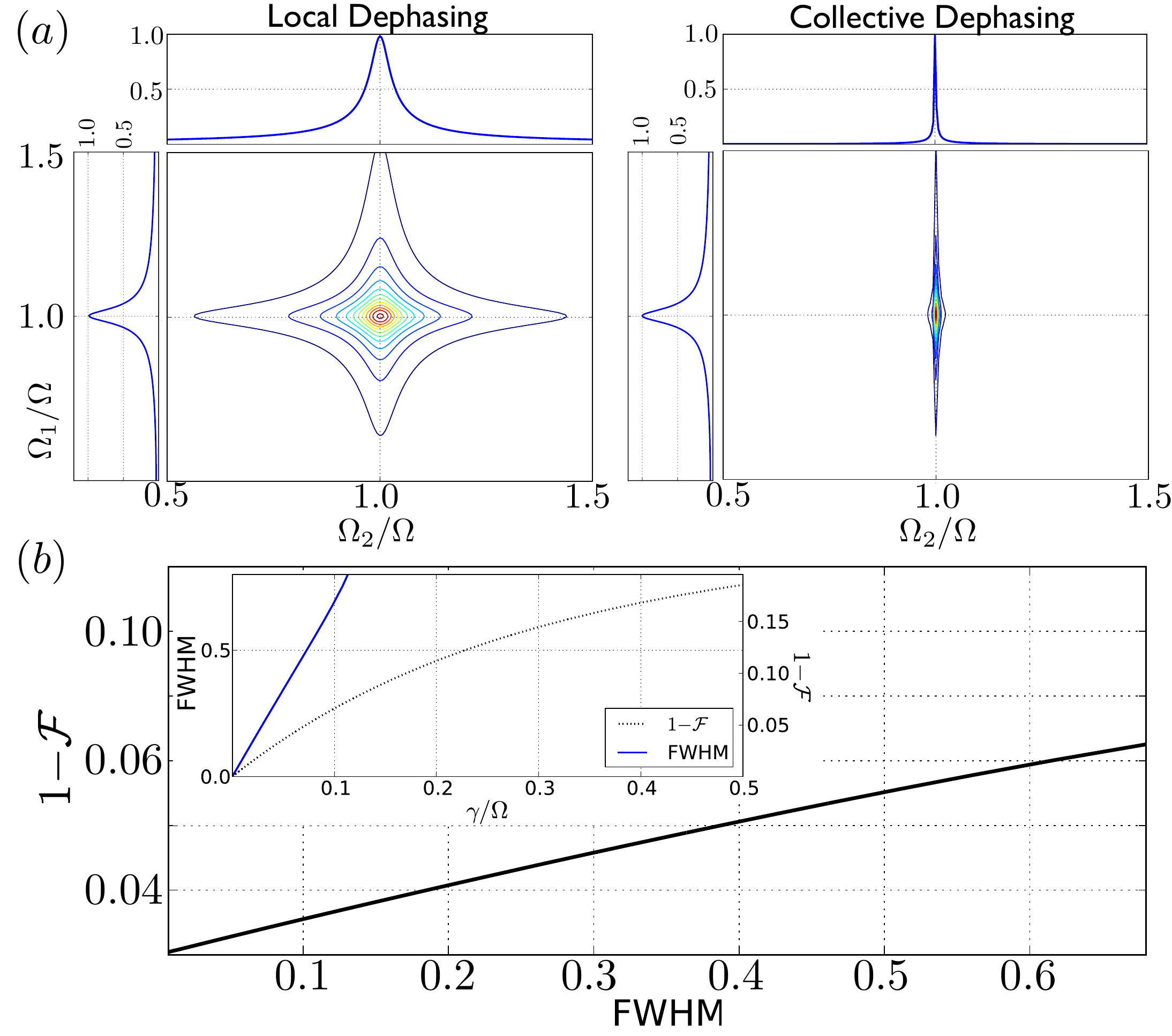}
\caption{Decoherence mechanisms can be identified and quantified via two-dimensional lineshapes. (a) $(\Omega,\Omega)$-signal of $|S^{\textit{(SQC)}}_{1,1;1} (\Omega_1, \Omega_2)|$ as read off from two spins initially prepared in the state $|00\rangle$ and subject to the M\o lmer-S\o rensen interaction. The two-dimensional lineshapes clearly distinguish between local and collective dephasing processes. Only under collective dephasing the system evolves in a decoherence-free subspace during $t_2$, as manifest in the delta-shaped signal along $\Omega_2$. (b) For weak local dephasing ($\gamma<0.1\Omega$), the full-width-half-maximum (FWHM) along $\Omega_1$ scales approximately linearly with the corresponding M\o lmer-S\o rensen gate error $1-\mathcal{F}$, since both relate to the strength of the noise $\gamma$ (inset).}
\label{fig.spins}
\end{figure}

The formalism for the construction of nonlinear signals introduced before can be readily adapted to other degrees of freedom by identifying appropriate excitation and readout schemes. For spins, we consider the interaction $\mathcal{V}_{j}(\phi)\rho=U_j(\phi)\rho U^{\dagger}_j(\phi)$, where $U_j(\phi)=\alpha\mathbb{I}^{(j)}+\beta(e^{i\phi}\sigma_+^{(j)}-e^{-i\phi}\sigma_-^{(j)})$, $|\alpha|^2+|\beta|^2=1$, can be generated by a focussed resonant laser pulse on spin $j$ \cite{Leibfried}, with $\sigma_{\pm}^{(j)}=(\sigma_{x}^{(j)}\pm i\sigma_{y}^{(j)})/2$, and $\sigma^{(j)}_{x,y,z}$ the Pauli matrices of spin $j=1,2$. We choose the readout observable $A_j=\sigma_z^{(j)}$, corresponding to a measurement of the population at spin $j$ \cite{Leibfried}.

The effective spin-spin interaction is described by the M\o lmer-S\o rensen Hamiltonian, which for two spins takes the form $H_{\mathrm{MS}}=(\Omega/2)\sigma^{(1)}_x\otimes\sigma^{(2)}_x$. The two ions are initialized in their electronic ground state $|00\rangle$, and all excitations and the readout are carried out on the same spin. One can readily predict four resonances at $(\Omega_1,\Omega_2)=\pm(0,\Omega)$ and $\pm(\Omega,\Omega)$, generated by coherences of Bell-states $\{|\Psi^{\pm}\rangle,|\Phi^{\pm}\rangle\}$, which are eigenstates of $H_{\mathrm{MS}}$ with eigenvalues $\pm\Omega/2$  \cite{MS}. Without dissipation, all four peaks are delta peaks, yet the coupling to an environment can broaden them and gives rise to distinct two-dimensional lineshapes. Figure~\ref{fig.spins} (a) compares the lineshape of the $(\Omega,\Omega)$-peak under the effect of a local noise process, described by the two Lindblad operators $L_{\mathrm{ld}}^{(j)}=\sqrt{\gamma}\sigma_z^{(j)}$, $j=1,2$, with the lineshape under correlated dephasing described by $L_{\mathrm{cd}}=\sqrt{\gamma}\sigma^{(1)}_z\otimes\sigma^{(2)}_z$. Local dephasing broadens the resonances equally along both frequency axes, whereas collective dephasing only affects the width along $\Omega_1$. The selected quantum pathway evolves in the coherence $|\Psi^{\pm}\rangle\langle\Psi^{\mp}|$ during $t_2$, which is part of a decoherence-free subspace \cite{Lidar}. For local dephasing, the peak width along $\Omega_1$ allows to infer the error probability of the corresponding gate $U_{\mathrm{MS}}=\exp{(-iH_{\mathrm{MS}}\pi/2 \Omega)}$, since both these
quantities scale approximately linearly with the dephasing strength $\gamma$ for $\gamma<0.1\Omega$ [Figure~\ref{fig.spins}(b)]. The error probability is quantified via the fidelity $\mathcal{F}=\sqrt{\langle 00|U_{\mathrm{MS}}^{\dagger}\rho_{\gamma}U_{\mathrm{MS}}|00\rangle}$ as $1-\mathcal{F}$, where $\rho_{\gamma}=\mathcal{G}(\pi/2\Omega)[|00\rangle\langle 00|]$ denotes the state at the outcome of the noisy gate.

\section{Conclusion}
We have presented a powerful method to probe spatially resolved multi-time correlation functions for the investigation of complex nonequilibrium dynamics and discussed experimental realizations with trapped ions. Our methods generalize previous Ramsey-type techniques inasmuch as they allow for the construction of pulse sequences of arbitrary order and phase-coherently select the contribution of individual quantum pathways.

Our diagrammatic theory provides an intuitive description as well as a new language for cold matter experiments and can be readily extended to include recent developments from nonlinear spectroscopy, such as pulse-shaping and optimal control theory \cite{control, Motzkus}. We envision that, beyond trapped ions, implementations with cold Rydberg gases \cite{Tweezer,Weidemueller} and spins of neutral atoms in optical lattices are within experimental reach \cite{Lukin13,Weitenberg11}.

We thereby provide a versatile toolbox for controlled quantum systems, opening up a wide range of possibilities to systematically study many-body effects in complex quantum systems, such as the survival of coherences under different environmental and internal couplings, as well as the role of excitonic states in quantum transport processes. The number of pulses which are required to probe a certain multi-point correlation function is independent of the system size, which renders this method scalable in the limit of increasing particles. This establishes an important step towards the experimental certification of quantum effects in large-scale quantum devices.

\ack
We would like to thank Michael Ramm for helpful discussions. M.G. and F.S. thank the German National Academic Foundation for support. S.M. gratefully acknowledges the support of the National Science Foundation through Grant No. CHE-1058791, and the Chemical Sciences, Geosciences and Biosciences Division, Office of Basic Energy Sciences, Office of Science, US Department of Energy. This work was supported by the NSF CAREER Program Grant No. PHY 0955650.

\appendix

\section{Fluorescence readout of phonon populations}\label{sec.A}
Populations of the vibrational degree of freedom can be probed using sideband transitions, i.e., by driving the electronic resonance with a laser detuning which corresponds to the trap frequency. Specifically, addressing the first red sideband (with a red detuning of $-\nu_x$) of an electronic transition between two states $\vert \uparrow \rangle$ and $\vert \downarrow\rangle$ with a pulse duration $t=\pi/\eta\Omega$ maps an initial state $|\downarrow,n\rangle$ to
\begin{equation}
U_{-1}(\pi)|\downarrow,n\rangle=\cos\left(\sqrt{n}\frac{\pi}{2}\right)|\downarrow,n\rangle+\sin\left(\sqrt{n}\frac{\pi}{2}\right)|\uparrow,n-1\rangle,
\end{equation}
with the Rabi frequency $\Omega$ of the carrier transition (resonant to the electronic levels) and we have neglected contributions of the order $\eta^2$, where $\eta$ denotes the Lamb-Dicke parameter \cite{Leibfried}. This pulse is followed by fluorescence readout of the electronic excitation, which can be done conveniently and with high efficiency using a short-lived excited electronic state which is coupled only to the electronic ground state, but not to the excited state \cite{Leibfried}. The probability to detect the ion in the excited state is then given by $|\sin(\sqrt{n}\pi/2)|^2$ and this measurement sequence is thus equivalent to effectively measuring the motional observable
\begin{equation}
A=\sum_{n=0}^{\infty}\sin^2\left(\sqrt{n}\frac{\pi}{2}\right)|n\rangle\langle n|.
\end{equation}
This sequence can be applied to single ions in a tight laser focus of sufficient intensity, to probe local vibrational excitations.

\section{Nonlinear spectroscopy and phase cycling}\label{sec.pc}
To obtain multidimensional spectra one applies a series of pulses with well-defined phase relations and tunable time-delays to the system of interest. The signal is measured as a function of the time-delays between the pulses. Considering, for instance, an excitation of the form $D_j(\alpha e^{i\phi})\approx \mathbb{I} + \alpha e^{i\phi}a_j^{\dagger} - \alpha e^{-i\phi}a_j$, the created coherences will carry the phase-shift $e^{\pm i\phi}$. All the contributions in a sequence of pulses can be represented by Feynman ladder diagrams, each one representing one excitation pathway with a characteristic phase signature (the combined phase shift of all pulses in the applied sequence), as for instance in Fig.~\ref{fig.2} a) and Fig.~\ref{fig.3} a). The diagrams are to be read as follows: 
\begin{itemize}
\item Time runs from bottom to top, the left vertical depicts the evolution of the \textit{ket}, and the right one the evolution of the \textit{bra} side of the density matrix.
\item Each excitation (de-excitation) is described by an arrow pointing towards (away from) the density matrix. Each de-excitation adds a factor $(-1)$ to the overall sign of the diagram.
\item To yield a signal, the diagram has to end up in an excited state population, when the fluorescence is collected.
\end{itemize}
The total signal is described by a coherent superposition of all pathways. A complete pedagogical introduction to this formalism is beyond the scope of the current paper, but can be found in a recent review \cite{ScholesReview}, which is specifically written to appeal to quantum opticians, as well as in standard textbooks \cite{HammZanni,Shaul_book}.

Phase cycling is a post-processing method which allows us to extract the contribution of subsets of pathways from the total signal, by exploiting their dependence on the phases \cite{Ernst}. To this end, the phases $\phi_i$ of the individual pulses have to be scanned over a discrete set of values.

As an example we consider a pulse sequence, where the phase signatures $k\Delta\phi$, $k=0,\pm1,\dots,\pm k_{\mathrm{max}}$ occur, where in a perturbative treatment of the low-intensity pulses as considered above, $k_{\mathrm{max}}$ can typically be assumed small (i.e. $k_{\mathrm{max}}=1$ or $2$). From the total signal, which decomposes into a sum of terms with different phase shifts,
\begin{equation}
S^{(2)}(\Delta\phi)=\sum_{k=-k_{\mathrm{max}}}^{k_{\mathrm{max}}}S^{(2)}_ke^{ik\Delta\phi},
\end{equation}
we can extract the complex-valued terms $S^{(2)}_k$ using an inverse discrete Fourier transform:
\begin{equation}
S^{(2)}_k=\frac{1}{2k_{\mathrm{max}}+1}\sum_{j=0}^{2k_{\mathrm{max}}}S^{(2)}(\delta\phi_j)e^{-i k\delta\phi_j},
\end{equation}
with $\delta\phi_j=2\pi j/(2k_{\mathrm{max}}+1)$. Thus, by scanning the phase shifts over a specifically chosen set, followed by an inverse discrete Fourier analysis of the obtained spectra, it is possible to experimentally select the contribution of individual pathways.

\end{document}